# Molecular simulation of translational and rotational diffusion of Janus nanoparticles at liquid interfaces

Hossein Rezvantalab[1], German Drazer[1], and Shahab Shojaei-Zadeh[1,2,*]

[1]*Department of Mechanical and Aerospace Engineering, Rutgers, The state University of New Jersey, 98 Brett Road, Piscataway, New Jersey 08854-8058, United States*

[2]*Institute for Advanced Materials, Devices and Nanotechnology, 607 Taylor Road, Piscataway, New Jersey 08854-8019, United States*

[*]*Corresponding author, E-mail: zadeh@rutgers.edu*

**Abstract**

We perform molecular dynamics simulations to understand the translational and rotational diffusion of Janus nanoparticles at the interface between two immiscible fluids. Considering spherical particles with different affinity to fluid phases, both their dynamics as well as the fluid structure around them are evaluated as a function of particle size, amphiphilicity, fluid density, and interfacial tension. We show that as the particle amphiphilicity increases due to enhanced wetting of each side with its favorite fluid, the rotational thermal motion decreases. Moreover, the in-plane diffusion of nanoparticles at the interface becomes slower for more amphiphilic particles, mainly due to formation of a denser adsorption layer. The particles induce an ordered structure in the surrounding fluid that becomes more pronounced for highly amphiphilic nanoparticles, leading to increased resistance against nanoparticle motion. A similar phenomenon is observed for homogeneous particles diffusing in bulk upon increasing their wettability. Our findings can provide fundamental insight into the dynamics of drugs and protein molecules with anisotropic surface properties at biological interfaces including cell membranes.

## I. INTRODUCTION

The adsorption of nanometer-sized objects, including nanoparticles, dendrimers and proteins, at soft interfaces has recently attracted much scientific interest and is central to a number of emerging technologies.[1-3] Self-assembly of nanoparticles at fluid interfaces has enabled the preparation of high quality two-dimensional crystals that can be employed for the fabrication of capsules, ultra-thin cross-linked membranes, and free-standing metal films.[4-6] The modification of interfacial properties by the adsorption of nanoparticles may be used to stabilize micrometer-scale structures such as nanoparticle-armored fluid droplets or phase-arrested gels.[7, 8] Closely related to the behavior of synthetic nanoparticles at interfaces are the interactions of biological macromolecules, proteins and virus capsids with liquid interfaces and cell membranes.[9-17]

While the adsorption of nanoparticles with homogeneous surface properties to liquid interfaces can represent several industrial and biological processes, the advent of particles with anisotropic surface characteristics has opened up a variety of emerging applications. The term "Janus" spans a wide class of particles composed of two chemically or physically distinctive surfaces.[18] Efficient high-throughput methods have been developed for fabrication of such particles with various shapes *e.g.* spheres, ellipsoids, cylinders and disks.[19-21] Advances in manufacturing processes of Janus particles have substantially widened the range of possible surface properties and resulting applications. For example, electrically anisotropic Janus particles are potentially useful in electronic paper and switchable display panels,[22-24] magnetic ones can be used as microrheological probes or biosensors,[25, 26] and those with amphiphilic surface properties can act as particulate surfactants at liquid interfaces.[27, 28] In the latter type, the wettability of each side can be tuned by grafting appropriate hydrophilic/hydrophobic groups on the particle surface. The anisotropy in wetting properties of such particles leads to unique characteristics that are not observed in homogeneous particles of similar dimensions. For instance, they exhibit much stronger binding to liquid interfaces, thus promoting the stability of particle-coated interfaces (foams and emulsions) which play an important role in a variety of applications such as in crude oil recovery, cosmetics, drug delivery, and food preservation.[29-34]

The amphiphilicity of Janus particles can also be exploited in chemical reactions, *e.g.* in enhancing the phase-selectivity of biofuel refining processes.[35] Moreover, bio-compatible Janus nanoparticles can be a promising new class of drug and gene delivery agents as well as agents for targeted destruction of cancer cells.[14, 36, 37] Nanoparticle properties, such as their size, shape, and surface chemistry influence their interactions with cell membranes.[38-40] A detailed description of these interactions at a molecular level is essential for designing nanoparticles with programmable and controlled membrane translocation mechanisms. Understanding the stability and interactions of Janus nanoparticles at liquid interfaces can serve as a platform for predicting particle-membrane interactions.

Molecular simulations to date have revealed important aspects of the physical behavior of nanoparticles both in bulk fluids as well as at an interface. For example, the stability and desorption energy of Janus nanoparticles from a liquid interface were recently investigated.[41, 42] Additionally, contact angle measurements have been performed on silica nanoparticles with different distribution of functional groups at an oil-water interface, revealing that the contact angle of a nanoparticle with randomly dispersed hydrophilic/hydrophobic groups is smaller than that with equal mean surface chemistry composition but with each functional group concentrated only on one hemisphere.[43] On the other hand, the diffusion properties and rotational relaxation of such particles at a liquid interface have yet to be quantified. Although such analysis is available for homogeneous particles,[44] the effect of surface heterogeneity on tuning the diffusivity and rotational motion has not been investigated. The translational motion of Janus nanoparticles at liquid interfaces is important to understand interfacial self-assembly processes and to use them as tracer elements in micro- and nanorheology,[45, 46] while their orientational motion has implications for their use as catalysts *e.g.* at oil-water interfaces.[47] It is thus important to investigate the effect that heterogeneous surface chemistry has on the thermal motion of Janus particles at fluid interfaces.

In this study, we use molecular dynamics simulations to quantify the diffusivity of nano-sized Janus particles at the interface between two immiscible fluids. We consider spherical clusters of solid atoms with differing affinity to the fluid phases, and evaluate their dynamics and induced fluid structure as a function of particle size, amphiphilicity, fluid density, and interfacial tension. Molecular simulations provide access to the structure of the fluid around the particles, which explains the observed changes in diffusion coefficient with the degree of amphiphilicity. Due to the increasing use of Janus nanoparticles as stabilizing agents, drug carriers, and interfacial catalysts, our analysis can be of great scientific and technological importance for understanding the basics of their transport and thermal motion at liquid interfaces.

## II. SIMULATION METHODOLOGY

### A. Model and computational scheme

We use Molecular Dynamics (MD) to investigate the diffusion of Janus nanoparticles at liquid interfaces. Due to the ability of molecular simulations to directly access the lengthscales relevant to nanoparticles, they provide a natural tool for the investigation of such systems. One of the advantages of Molecular Dynamics over continuum approaches is that it provides information on fluid structure around the particles. At a liquid interface, the effects of line tension and capillary waves on the detachment energy of the particle can be evaluated, while these are neglected in the continuum approach which assumes a sharp circular contact line between the particle and the fluids.[48] These effects can significantly change particle motion at the interface for nano-sized particles. While the continuum model predicts that the equilibrium orientation of Janus spheres results in each hemisphere being exposed to its more favorable fluid,[49] molecular simulations suggest that the former approach overestimates the particle detachment energy as a result of the restricting assumptions mentioned above.[41] As a consequence, Janus particles may adsorb to liquid interfaces with random (tilted) orientations, as suggested by some experiments.[50, 51]

A schematic representation of the geometry under consideration is shown in Fig. 1. Each individual fluid is modeled as a monotonic Lennard-Jones fluid,[52] and the domain is bounded by two stationary solid walls parallel to the interface plane. The particle contains atoms of two different species, denoted as *J1* and *J2* and moves as a rigid body. The rotational evolution of the particle is monitored by tracking the angle between the Janus boundary and the interface plane. The particle is initially placed at the interface with a specific orientation, and it can freely rotate or transfer to a bulk fluid due to thermal fluctuations. We evolve the particle position and orientation as well as its linear and angular velocity over time through evolution of imposed forces and torques.

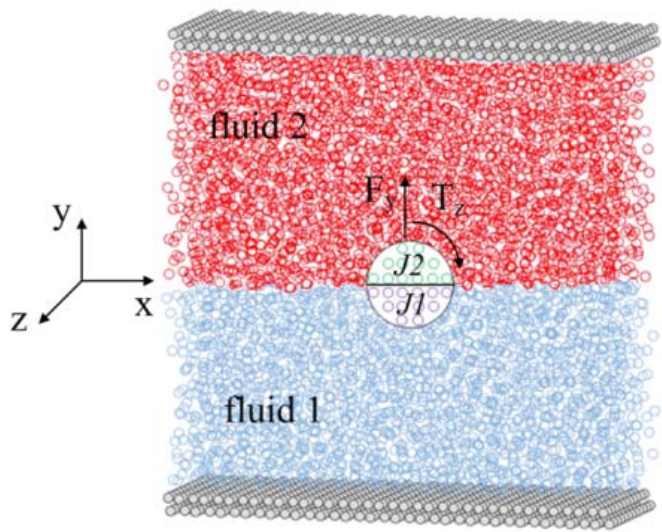

FIG. 1. Schematic representation of the simulation box containing a spherical Janus particle at the interface between two immiscible fluids.

Molecular Dynamics simulations are carried out using the shifted-force Lennard-Jones (LJ) potential given by

$$U(r_{ij}) = \begin{cases} 4\varepsilon\left[\left(\frac{\sigma}{r_{ij}}\right)^{12} - A\left(\frac{\sigma}{r_{ij}}\right)^{6}\right] - U(r_c) - \left.\frac{dU}{dr}\right|_{r_c}(r - r_c) & r \leq r_c \\ 0 & r > r_c, \end{cases} \qquad (1)$$

in which $r_{ij}$ is the separation distance between atoms $i$ and $j$, $\sigma$ is roughly the size of the repulsive core, $\varepsilon$ is the depth of the potential well, and $r_c = 2.5\sigma$ is the potential cutoff distance. The coefficient $A$ controls the attraction between solid/fluid atomic species in the system, which in turn defines the surface tension and wetting properties. The interaction coefficient for atoms in the same fluid and with the bounding walls are set to $A_{pp} = A_{pw} = 1$ (with $p = f1, f2$), while atoms of the two different fluids interact with $A_{f1f2} = 0.5$ in order to ascertain the immiscibility of the fluids. For the particle-fluid interaction, the partially wetting situation corresponds to values of $A$ less than 1. This parameter can be correlated to the contact angle $\theta$ at a solid–liquid interface approximately by $cos\theta = -1 + 2A$.[53, 54]

In this study, we consider Janus nanoparticles with two regions of opposite wettability (*i.e.*, apolar and polar), represented by $\theta_a = 90°+\beta/2$ and $\theta_p = 90°-\beta/2$, with the parameter $\beta = \theta_a - \theta_p$ characterizing the degree of amphiphilicity. This type of symmetrically-Janus particles has been also used in other relevant works considering Janus spheres and ellipsoids.[49, 55-58] Nevertheless, our analysis will generally hold for Janus particles comprised of any two polar and apolar regions. We increase the amphiphilicity by increasing the attraction coefficients between each side and its favorite fluid $A_{f1J1}, A_{f2J2}$, while symmetrically reducing the attraction with the opposite fluid $A_{f1J2}, A_{f2J1}$ (see Fig. 1). In this way, tuning the amphiphilicity in the range of $\beta = 0-180°$ (from homogeneous to most amphiphilic) corresponds to increasing the attraction coefficients $A_{f1J1}, A_{f2J2}$ from 0.5 to 1, while reducing the cross-interactions from 0.5 to 0. More details of the simulation procedure are given in Appendix A.

## B. Analysis

For nanoparticles diffusing in bulk fluid, the translational motion can be characterized by the diffusion coefficient

$$D_t = \frac{1}{6}\frac{d}{dt}\langle \Delta r_p(t)^2 \rangle = \frac{1}{6}\frac{d}{dt}\langle (r_p(t) - r_p(0))^2 \rangle, \qquad (2)$$

where $r_p(t)$ is the center of mass position of the particle at time $t$ and the angle brackets denote averaging over realizations with different initialization of fluid velocity. On the other hand, the rotational motion of the nanoparticle can be quantified in terms of the angular velocity correlation function

$$C_\omega(t) = \langle \boldsymbol{\omega}(t) \cdot \boldsymbol{\omega}(0) \rangle, \qquad (3)$$

where $\boldsymbol{\omega}(t)$ is the angular velocity vector at time $t$. The rotational diffusion coefficient can be calculated by integrating the correlation function

$$D_r = \frac{k_B T}{I} \int_0^\infty \frac{C_\omega(t)}{C_\omega(t=0)} dt, \qquad (4)$$

where $I$ is the mass moment of inertia of the particle which is determined by adding the moment of each atom within the cluster about the center of mass.

In case of immiscible fluids at a liquid interface, the nanoparticle diffusion becomes anisotropic and it is thus useful to decompose the motion of the nanoparticle into parallel and normal components, as shown schematically in Fig. 2(a). Specifically, we calculate the mean-squared displacement within the interface plane (*xz* plane) and normal to it (in *y* direction), and the diffusivity can then be obtained from the time derivatives as

$$D_t^{\parallel} = \frac{1}{4dt}\langle \Delta r_p^{\parallel}(t)^2 \rangle = \frac{1}{4dt}\langle (x(t)-x(0))^2 + (z(t)-z(0))^2 \rangle$$
$$D_t^{\perp} = \frac{1}{2dt}\langle \Delta r_p^{\perp}(t)^2 \rangle = \frac{1}{2dt}\langle (y(t)-y(0))^2 \rangle.$$
(5)

The motion normal to the interface is generally bounded, leading to a finite MSD (zero long time diffusivity). As an alternative to the mean-squared displacement approach, the diffusion coefficient can be quantified using the velocity autocorrelation function $C_v^{\parallel}(t) = \langle v^{\parallel}(t) \cdot v^{\parallel}(0) \rangle$ for the motion parallel to the interface[59]

$$D_t^{\parallel} = \frac{1}{2}\int_0^{\infty} C_v^{\parallel}(t)\, dt.$$
(6)

We use both formulations to show their consistency for a sample case, but the latter approach is employed for the rest of this study. In terms of particle rotation, homogeneous nanoparticles diffuse isotropically around the three coordinate axes. Janus particles, on the other hand, can freely rotate around the direction normal to the interface, but rotation is limited in the other directions as the contact line becomes locked at the Janus boundary. The nanoparticle rotation can be decomposed into in-plane (around *y* axis) and out-of-plane (around *x,z* axes) as

$$D_r^{\parallel} = \frac{k_B T}{I}\int_0^{\infty} \frac{C_\omega^n(t)}{C_\omega^n(t=0)} = \frac{k_B T}{I}\int_0^{\infty} \frac{\langle \omega_y(t) \cdot \omega_y(0) \rangle}{\langle \omega_y(0) \cdot \omega_y(0) \rangle}$$
$$D_r^{\perp} = \frac{k_B T}{I}\int_0^{\infty} \frac{C_\omega^p(t)}{C_\omega^p(t=0)} = \frac{k_B T}{I}\int_0^{\infty} \frac{\langle \omega_x(t) \cdot \omega_x(0) + \omega_z(t) \cdot \omega_z(0) \rangle}{\langle \omega_x(0) \cdot \omega_x(0) + \omega_z(0) \cdot \omega_z(0) \rangle}.$$
(7)

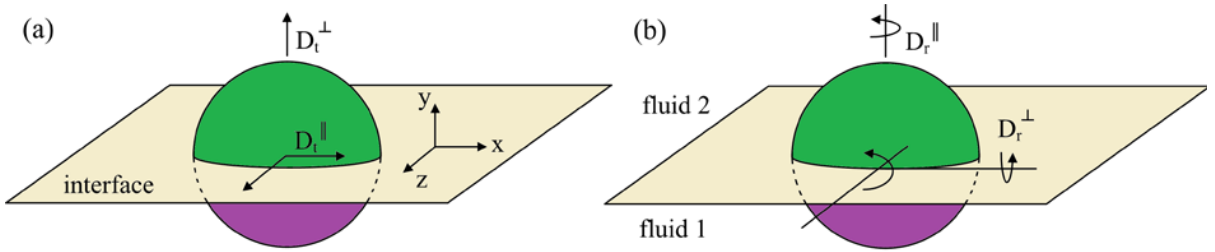

FIG. 2. Schematic view of the in-plane and out-of-plane decomposition of (a) translational and, (b) rotational diffusion of a Janus particle at a liquid interface.

This is shown schematically in Fig. 2(b). We will thus compare the in-plane component of the rotational diffusion of nanoparticles with different surface properties since $D_r^\perp$ does not demonstrate a diffusive behavior for Janus particles due to stronger binding of each hemisphere with its favorite fluid. Alternatively, the diffusion can be quantified considering the in-plane rotation of a unit vector attached to the particle through the orientational correlation function

$$C_e(t) = \langle \hat{\boldsymbol{e}}(t) \cdot \hat{\boldsymbol{e}}(0) \rangle, \tag{8}$$

where $\hat{\boldsymbol{e}}(t)$ is taken as the unit vector connecting two atoms in the particle and is initially oriented along x direction. The rotational diffusion coefficient can be estimated from the slope of this exponentially decaying function at long times[59, 60]

$$D_r^\parallel = -\frac{1}{2} \lim_{t \to \infty} \frac{d}{dt} \ln[C_e(t)]. \tag{9}$$

Similar to translational diffusion, we employ the method based on integration of the angular velocity correlation function $C_\omega(t)$, unless otherwise stated.

## III. RESULTS AND DISCUSSION

We first evaluated the average fluid properties in the absence of nanoparticles. The numerical method for calculating the fluid density, diffusivity, and the interfacial tension between the two phases, as well as their estimated values are presented in Appendix B. The results indicate excellent agreement with theoretical correlations and earlier simulations. Furthermore, the diffusion of homogeneous nanoparticles in bulk fluid was investigated for several particle sizes in order to validate the simulation method. The calculated translational and rotational diffusivities were compared with theoretical values based on Stokes-Einstein theory,[61] showing less than 6% deviation (see Appendix C).

## A. Janus nanoparticles at liquid interfaces

When the nanoparticle is adsorbed at a liquid–liquid interface, the diffusion becomes anisotropic since it is energetically favorable for the particle to remain adsorbed to the interface. The detachment energy required to remove the particle from an interface is given by[62]

$$\Delta E = \gamma \pi R^2 (1 + cos\theta)^2, \tag{10}$$

where $\gamma$ is the interfacial tension between the two fluids, and $\theta$ is the three-phase contact angle. For the homogeneous nanoparticles considered in this work, the contact angle is $\theta = 90°$ due to the symmetry in wetting coefficients with the two fluids. For Janus nanoparticles, each side has a higher affinity to one of the fluid phases, so that the particle is again adsorbed symmetrically relative to the interface.[49] In reduced units, we have $\sigma^2\gamma/k_BT = 0.59$ and setting $\theta = 90°$ yields normalized detachment energies of $\Delta E/k_BT = 16.6\text{-}66.5$ for the investigated size range. It has been shown that Eq. (10) is likely to underestimate the detachment energy due to neglecting the line tension and capillary wave effects.[48] Therefore, the detachment energy is several times larger than the thermal energy, such that the motion normal to the interface is expected to be significantly smaller than the in-plane diffusion, and the probability of spontaneous detachment is negligible. The desorption may only occur for fluids with significantly lower interfacial tensions which are not commonly encountered in stabilization or self-assembly purposes using such nanoparticles.

The mean-squared displacements for a Janus nanoparticle with $R = 3.0\ \sigma$ and an amphiphilicity characterized by $\beta = 120°$ is presented in Fig. 3(a) over the period $0 \leq t/\tau \leq 200$. We clearly note that the in-plane motion shows a linear diffusive trend, while the long time motion normal to the interface is non-diffusive. Our simulations indicate that the attachment becomes stronger as the particle amphiphilicity increases, due to the increased affinity between each hemisphere and its surrounding fluid.

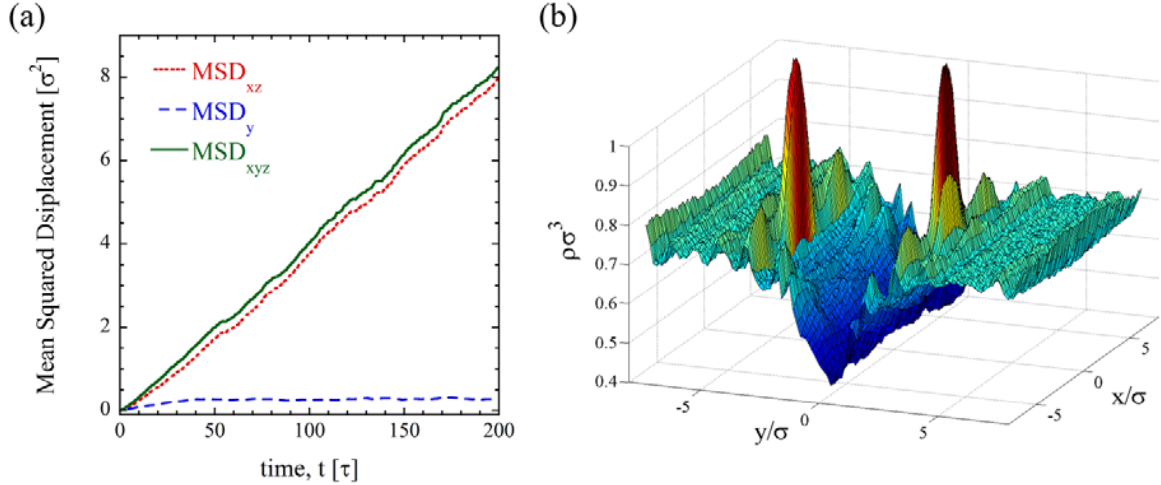

FIG. 3. (a) Mean-Squared Displacements for a Janus nanoparticle with size R = 3.0σ and amphiphilicity of β = 120° adsorbed at a liquid interface, (b) surface plot of fluid density around the particle.

Moreover, the fluid structure around the nanoparticle becomes anisotropic upon adsorption to the interface. An isometric view of the surface plot of fluid density variation in x-y plane is depicted in Fig. 3(b). Firstly, a depletion region is clearly detected between the two fluid components near $y = 0$ (in accordance with Fig. 10). Note that the size of the nanoparticles investigated here is comparable to the width of the depletion region at the interface.

More importantly, we observe a structured fluid with distinct peaks and dips close to the particle surface. The highest peaks occur at $y \approx \pm 3.5\ \sigma$, while the peak strength gradually declines when moving away from the center of the box in x,y directions as the distance from the nanoparticle surface increases. The nanoparticle thus perturbs the fluid as reflected in the high-density layers near the surface. Such layering is basically similar to the case of homogeneous nanoparticles at a liquid interface.[44]

In order to evaluate the effect of surface wettability on the diffusivity of nanoparticles, molecular simulations are performed on particles with different amphiphilicity as characterized by $β = 0$-

TABLE I. In-plane diffusivities for nanoparticles of size R = 3.0σ and different amphiphilicity at a liquid interface.

|  | $D_t^{\parallel}$ (MSD) | $D_t^{\parallel}$ ($C_v$) | $D_r^{\parallel}$ ($C_e$) | $D_r^{\parallel}$ ($C_\omega$) |
|---|---|---|---|---|
| *β = 0°* | $(10.6 \pm 0.6) \times 10^{-3}$ | $(10.2 \pm 0.5) \times 10^{-3}$ | $(15.4 \pm 0.9) \times 10^{-4}$ | $(16.0 \pm 0.8) \times 10^{-4}$ |
| *β = 60°* | $(9.6 \pm 0.5) \times 10^{-3}$ | $(9.3 \pm 0.5) \times 10^{-3}$ | $(13.6 \pm 0.8) \times 10^{-4}$ | $(13.0 \pm 0.8) \times 10^{-4}$ |
| *β = 120°* | $(9.3 \pm 0.5) \times 10^{-3}$ | $(9.1 \pm 0.5) \times 10^{-3}$ | $(10.5 \pm 0.8) \times 10^{-4}$ | $(10.0 \pm 0.8) \times 10^{-4}$ |
| *β = 180°* | $(9.1 \pm 0.5) \times 10^{-3}$ | $(9.0 \pm 0.5) \times 10^{-3}$ | $(9.8 \pm 0.8) \times 10^{-4}$ | $(9.6 \pm 0.8) \times 10^{-4}$ |

*180°*. The results for *R = 3.0 σ* are shown in Table I along with the associated uncertainty, indicating a good agreement between the values found using different methods. The difference generally lies within the statistical uncertainty in each case. We observe that both the in-plane translational and rotational diffusion coefficients are reduced as the amphiphilicity of the nanoparticles increases.

In order to identify the source of the observed decrease in $D_t^{\parallel}$ for particles of increasing amphiphilicity, we look at the molecular structure of the fluid surrounding these nanoparticles. The radial distribution function *g(r)* for Janus nanoparticles of different amphiphilicity is shown in Fig. 4. We observe that in all cases, the fluid is highly structured around the particle, with a number of peaks spaced by ~*1σ*. More importantly, the fluid layering becomes more pronounced as *β* increases. This microscopic observation indicates that the fluid structures more uniformly into layers around amphiphilic Janus particles compared to a homogeneous one with the same size but *90°* contact angle.

As the amphiphilicity increases, the wetting of each hemisphere with its surrounding fluid is enhanced. Earlier molecular simulations for water near a solid wall suggest that water molecules adjacent to more hydrophobic walls are more randomly distributed and show a weaker ordering.[63] We observe a similar behavior for amphiphilic spheres, with a stronger ordering as each side is more preferably wetted by the surrounding fluid. We expect such increase in layering to induce a larger resistance against particle motion as the adsorption layer of fluid atoms around the particle will contain a higher concentration of fluid atoms. Consequently, nanoparticles with higher amphiphilicity experience a slower diffusion within the interface plane. In addition, the fluctuation of the contact line around the Janus boundary and the exposure of solid atoms to unfavorable fluid species may also contribute to a reduction in diffusivity and thus slightly couple particle translation to its rotation. To evaluate this hypothesis, we repeated our simulations but with the added constraint of fixed upright orientation for Janus nanoparticles of $R = 3.0\ \sigma$. The result demonstrates that the decrease in the in-plane translational diffusion coefficients with amphiphilicity reduces, while the fluid structure is almost unaffected. For instance, the decrease in $D_t^\parallel$ between particles with $\beta = 0°$ and $180°$ is reduced from 14% to ~8% upon fixing the orientation of the particles. Therefore, we conclude that the in-plane diffusion slows down with increased amphiphilicity due to both stronger fluid layering as well as finite fluctuations around the upright orientation.

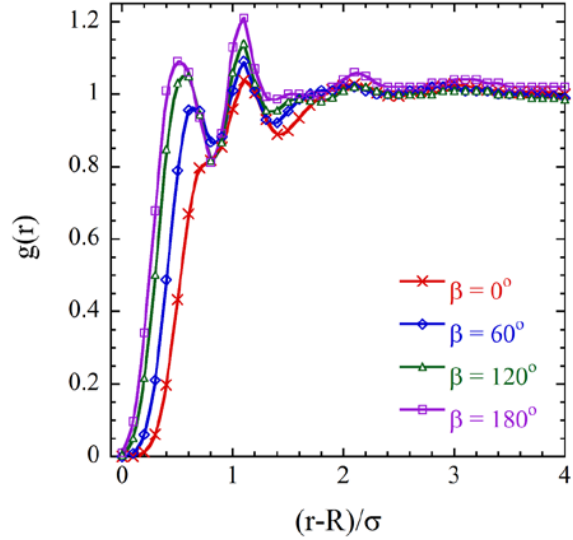

FIG. 4. Radial distribution function for the fluid surrounding nanoparticles with R = 3.0σ and different amphiphilicity at a liquid interface.

For the rotational diffusion, the slower in-plane rotation of more amphiphilic nanoparticles can also be attributed to the increased resistance of the adsorbed fluid layer. Therefore, particles of similar sizes but different surface properties experience different translational and rotational diffusion at a liquid interface.

In order to further clarify the role of fluid layering on the diffusivity, homogeneous nanoparticles with wettabilities corresponding to favorable solid/fluid interactions in the above Janus particles were simulated in the bulk. The result indicates a stronger fluid layering upon increasing the wettability, followed by a slower diffusion. This confirms that the increased affinity of each solid surface to its surrounding fluid for the Janus particles contributes to the reduction in diffusion coefficients. However, the reduction is generally larger in case of Janus nanoparticles at the interface.

### B. Role of nanoparticle size on their interfacial diffusivity

In general, larger particles exhibit a slower diffusion with longer rotational relaxation time.[44] In a bulk fluid, the translational and rotational diffusion coefficients are inversely proportional to $R$ and $R^3$, as predicted by Stokes-Einstein(-Debye) theory in Eqs. (C1,C2). The diffusion coefficients are however modified at a liquid interface due to the change in the effective drag acting on the nanoparticles. The presence of surface heterogeneity in Janus nanoparticles further complicates the issue as shown in the previous section. The variation of in-plane translational diffusion coefficient $D_t^{\parallel}$ with nanoparticle size and amphiphilicity is presented in Fig. 5(a). For all investigated particle sizes, we observe a reduction in diffusion coefficients upon increasing the amphiphilicity. Moreover, the diffusion becomes slower by increasing the nanoparticle size at each $\beta$, but our calculations indicate that the rate of decay is faster than the $R^{-1}$ proportionality in case of particles in a bulk fluid. A similar trend is observed in case of rotational diffusion as shown in Fig. 5(b), with the reduction rate being generally faster than $R^{-3}$ for different amphiphilicities. As the particle size increases compared to the interfacial width, more of its surface becomes in contact with the bulk fluid, so that the effect of the interfacial region is suppressed. Consequently, the difference between interfacial and bulk diffusivities reduces, so that $D_t^{\parallel}$ and $D_r^{\parallel}$ drop with a faster slope with particle size as compared to the Stokes-Einstein(-Debye) prediction.

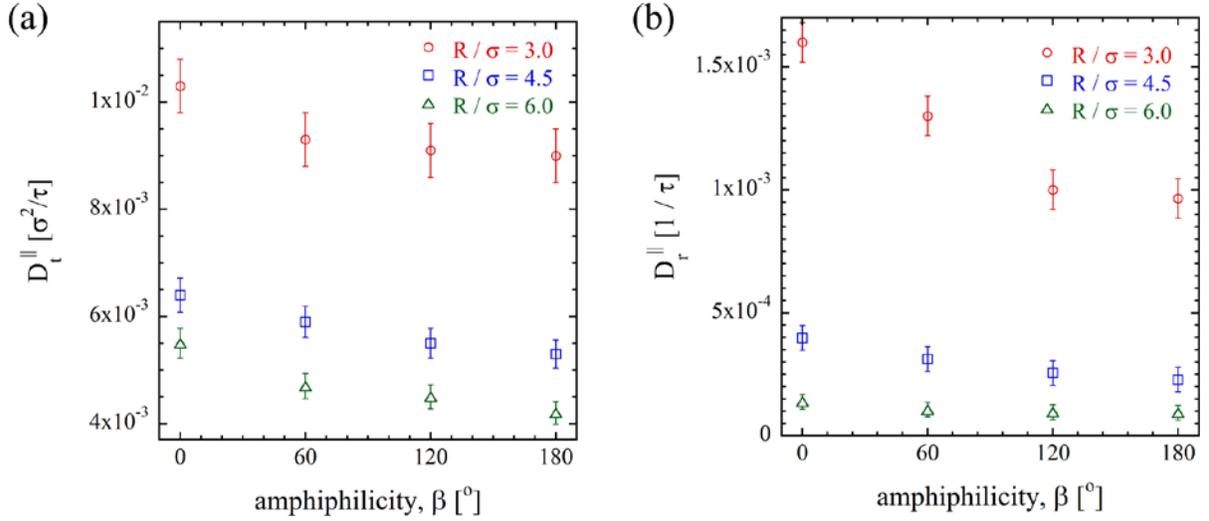

FIG. 5. Schematic (a) Translational and, (b) rotational diffusion coefficients for Janus nanoparticles of different size and amphiphilicity at the interface between two immiscible fluids.

## C. Effect of fluid density on interfacial diffusion of Janus nanoparticles

We demonstrated in section III.A that the fluid layering around nanoparticles at the interface is enhanced for increasing amphiphilicity. As a consequence, the adsorbed fluid layer becomes denser, thus inducing a larger resistance against particle motion. However, the fluid structure near a solid surface generally depends on the bulk density, with denser fluids being more structured.[63] Therefore, it is important to evaluate the effect of fluid density on layering around nanoparticles of different surface wettability. To do this, nanoparticles of $R = 3.0\,\sigma$ and different amphiphilicities characterized by $\beta = 0\text{-}180°$ are considered, while increasing/decreasing the density of surrounding fluids from the original value of $\rho = 0.8/\sigma^3$. Figure 6(a) shows the radial distribution function around nanoparticles for the case of $\rho = 0.7/\sigma^3$. We observe a trend similar to that of the denser fluid studied so far but with the first peak being considerably lower. This indicates that a lower-density fluid is generally less structured around the nanoparticles as a smaller number of fluid atoms can form an adsorption layer adjacent to the solid surface. As the fluid density is increased to $0.9/\sigma^3$, the first peak grows higher and the fluid becomes more structured at close vicinity of the particle as shown in Fig. 6(b). In addition, the peaks shift to the left and their spacing is reduced, suggesting more concentrated fluid atoms. For both fluid densities, we observe that the fluid ordering is enhanced as the amphiphilicity increases. This effect is more pronounced at lower densities and becomes less visible for higher-density fluids. In addition, we evaluate the translational and rotational diffusion coefficients of nanoparticles at the interface between these fluids. The calculated values for $D_t^{\parallel}$ are plotted in Fig. 7(a). The results indicate that the diffusivity increases as the fluid density is reduced. For Lennard-Jones fluids with the assigned temperature of $T = 1.0\ \varepsilon/k_B$ and a density of $\rho\sigma^3 = 0.7, 0.8, 0.9,$ the viscosity is measured as $\eta$ = *1.1, 2.0, 4.1 m.$\sigma^{-1}$.$\tau^{-1}$*, respectively.[64] Therefore, the enhanced diffusivity can be attributed to the reduction in the effective viscosity and drag force acting on the particle.

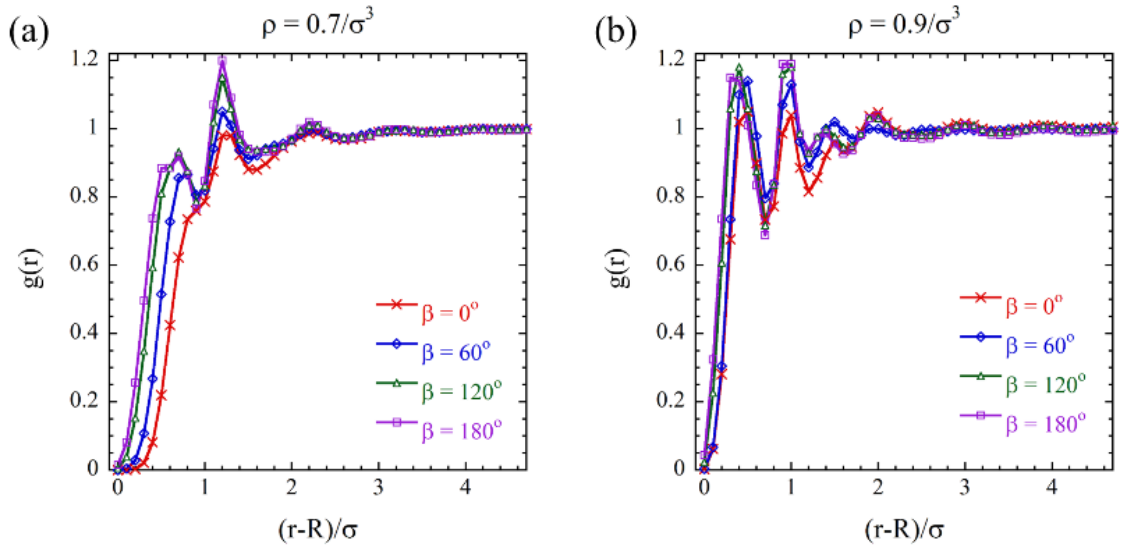

FIG. 6. (a) Radial distribution function for fluids with two different densities surrounding nanoparticles with R = $3.0\sigma$ at a liquid interface: (a) $\rho\sigma^3 = 0.7$, (b) $\rho\sigma^3 = 0.9$.

Equally important is understanding the role of wettability on interfacial diffusion. Figure 7(a) suggests that for a Janus particle with $\beta$ = *60°* compared to a homogeneous particle ($\beta$ = *0°*), the diffusivity reduces by ~*13, 10, 4%* at the interface between fluids with $\rho\sigma^3$ = *0.7, 0.8, 0.9*, respectively. Therefore, Janus nanoparticles of moderate amphiphilicity will practically diffuse similar to their homogeneous counterparts at the interface between two denser fluids. As shown in Fig. 6, denser fluids are generally more structured, so that the effect of nanoparticle surface wettability on fluid layering becomes less pronounced. On the other hand, the fluid layering around nanoparticles in lower density fluids is significantly affected by surface properties.

The fluid density and layering effects also influence the rotational thermal motion of nanoparticles at the interface. The in-plane rotational diffusivities for the above bulk densities are shown in Fig. 7(b), suggesting a trend similar to the translational diffusion. Particle rotation is reduced in denser fluids since the higher viscosity results in lager resistance against particle rotation. In addition, since the denser fluid has a stronger structure, this results in a more concentrated adsorption layer, which may be interpreted as an increased effective nanoparticle size and thus a slower translational and rotational diffusion.

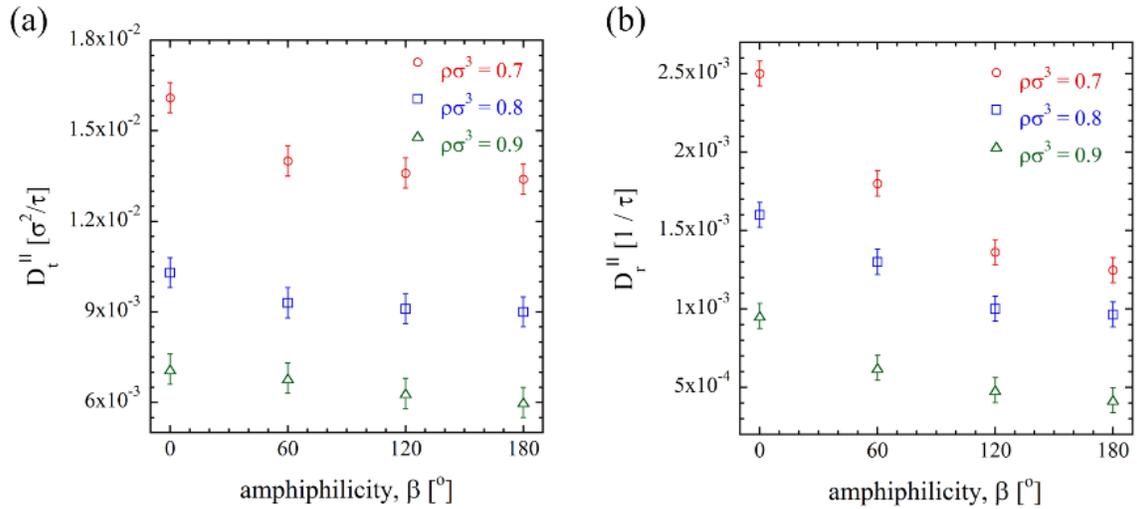

FIG. 7. (a) Translational, (b) rotational diffusion coefficients for nanoparticles of size R = 3σ and different amphiphilicity diffusing at the interface between fluids with various density.

D.  **The role of interfacial tension on diffusion of nanoparticles at the interface**

The interfacial motion of nanoparticles can be also influenced by the interfacial tension between the two fluid phases. We control the immiscibility of the two Lennard-Jones fluids through the attraction coefficient $A_{12}$, which was set to *0.5* in all previous simulations. On the other hand, the attraction coefficient between two atoms of the same fluid were set to $A_{11} = A_{22} = 1.0$. Increasing $A_{12}$ above *0.5* indicates a stronger attraction between the fluids yielding a lower interfacial tension. The change in interfacial tension will modify the density gap at the interface region and consequently the resistance against particle translation/rotation.

To evaluate this, we simulate the motion of nanoparticles of size $R = 3.0\ \sigma$ and amphiphilicities $0° \leq \beta \leq 180°$ at the interface between two fluids with density $\rho = 0.8/\sigma^3$, while varying the attraction coefficient between the two fluid species. In each case, the interfacial tension is estimated using Eq. (B5). The calculated values are shown in Table II, both in reduced MD units as well as converted into metric units using typical values for $\varepsilon$ and $\sigma$ given in Appendix A. We observe that for $A_{12} = 1.0$ where the two fluids are equivalent, there is no interfacial tension. As the attraction coefficient is reduced towards zero, the fluids become more immiscible, leading to an increased surface tension.

TABLE II. Calculated interfacial tensions for different attraction coefficients $A_{12}$ between two fluids in the simulation box.

| $A_{12}$ | 0.0 | 0.2 | 0.5 | 0.8 | 1.0 |
|---|---|---|---|---|---|
| $\gamma\ (\varepsilon/\sigma^2)$ | 2.46 | 2.13 | 1.47 | 0.36 | 0.00 |
| $\gamma\ (mN/m)$ | 45.3 | 39.2 | 27.0 | 6.62 | 0.0 |

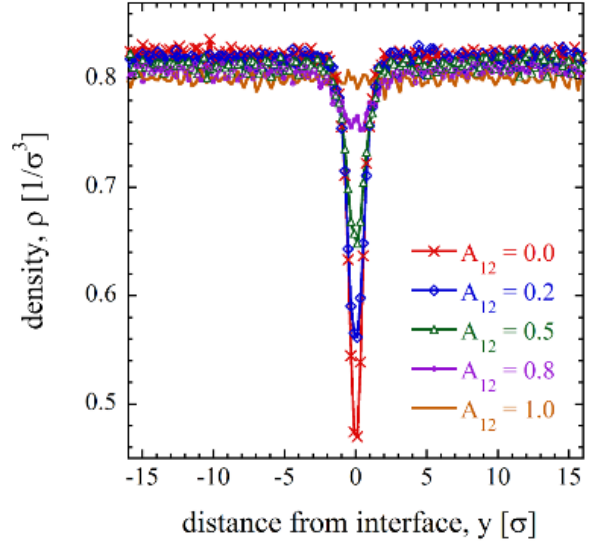

FIG. 8. Density profiles for two immiscible fluids as a function of distance from the interface for different attraction coefficients $A_{12}$ between the two fluids.

As shown in Fig. 8, the depletion region near the interface grows deeper as the attraction between the two species becomes weaker. For $A_{12} = 0$, the minimum density point is $\rho\sigma^3 \approx 0.47$. The increased repulsion between the two species clearly results in a smaller concentration of the fluid atoms near the interface.

The translational and rotational diffusion coefficients for nanoparticles of different amphiphilicities are shown in Fig. 9, revealing a slight increase in both as the surface tension increases. However, a much smaller variation in diffusion coefficients is detected compared to the effect of bulk density. As we change the surface tension, $D_t^{\parallel}$ and $D_r^{\parallel}$ deviate by ~11%, 9% on average.

The change in diffusivity with surface tension is probably due to deepening the density gap at the interface, which results in a slight reduction in effective viscosity and the drag force acting on the particle. The effect is much weaker than tuning the bulk fluid density though, since it only affects a small region with a width of $w \approx 2.5\sigma$ around nanoparticles. This effect diminishes even further for larger nanoparticles since the interfacial width becomes smaller relative to the particle size.

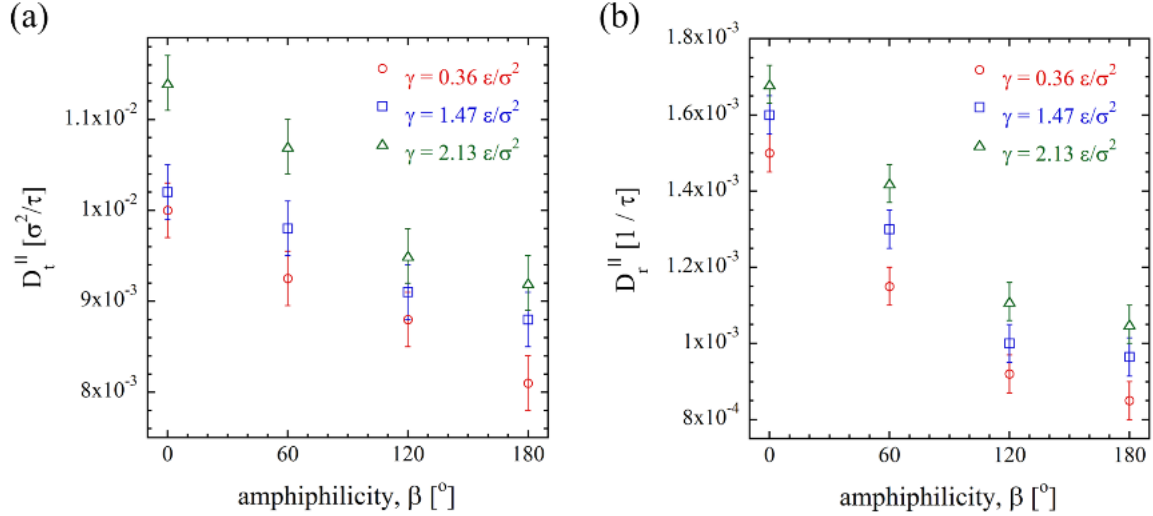

FIG. 9. (a) Translational, (b) rotational diffusion coefficients for NPs of size $R = 3\sigma$ and different amphiphilicity diffusing at liquid interfaces with various surface tensions.

**IV. CONCLUSIONS**

We have studied the diffusion of Janus nanoparticles at liquid–liquid interfaces using molecular dynamics simulations. The fluid atoms interact through a well-defined Lennard-Jones potential, while the immiscibility is controlled by the attraction between the two species. Janus nanoparticles are modeled as spherical clusters comprised of two sides with different affinity to fluid phases, with the attraction coefficient for each solid/liquid pair being related to the wettability. We demonstrated that the in-plane translational and rotational diffusion become slower upon increasing the amphiphilicity. By investigating the radial distribution function, we found that the fluid is more structured around Janus nanoparticles compared to their homogeneous counterparts. A larger number of fluid atoms contribute to the adsorption layer accompanying the particle, thus increasing the resistance against motion.

The effects of nanoparticle size, fluid density, and surface tension on the interfacial diffusivity were also investigated. We showed that both in-plane translational and rotational diffusion become slower with increasing particle size, but with decay rates being faster than those in a bulk fluid. Investigating the diffusion in fluids with different density revealed that the reduction in diffusion coefficient of Janus nanoparticles upon increasing the amphiphilicity can only be considerable in case of lower-density fluids, while the effect diminishes with increasing density. Furthermore, nanoparticles adsorbed at an interface with higher surface tension exhibit slightly slower diffusion. The results presented in this work provide insight into the dynamics of anisotropic nanoparticles at liquid interfaces and can have implications in evaluating the adhesion of drugs and protein molecules to cell membranes.

**ACKNOWLEDGEMENTS**

This research used resources of the National Energy Research Scientific Computing Center (NERSC), which is supported by the Office of Science of the U.S. Department of Energy under Contract No. DE-AC02-05CH11231. The authors are also grateful for support from Rutgers School of Engineering Computing Center.

**APPENDIX A. Simulation details**

The size of the simulation box is set to $L_x \times L_y \times L_z = 32\sigma \times 45\sigma \times 32\sigma$, and a size-independency test has been performed to ensure that the fluid properties are not affected by layering effects near the walls. Periodic boundary conditions are imposed in the directions parallel to the interface plane in order to simulate a large system and avoid wall effects.[65] The particle radius is set in the range of $R/\sigma = 3\text{-}6$ by cropping regions of different size from the initial cubic lattice.

The simulation starts with an equilibration process from the initial solid-like FCC lattice, followed by a post-equilibration stage for data collection. The positioning of different atomic species, inter-atomic forces, and kinetic and potential energies are measured over a sufficiently long period. The calculations are done in the canonical (NVT) ensemble, where the temperature is fixed using the Nosé-Hoover thermostat.[66, 67] The net force and torque on the particle are computed by adding up the individual interactions between its atoms and the neighboring fluid atoms, and the motion is governed by Newton's and Euler's equations. We use quaternion parameters to describe the particle orientation, in order to avoid the problem of divergence in the rotational equations of motion.[65] A Verlet list is built at frequent time-steps in order to minimize the calculation of interactions beyond the cutoff radius. Integrating Newton's and Euler's equations is done via a five-value predictor-corrector algorithm.

The results are presented in terms of dimensionless variables with the scaling parameters being $\sigma$ (length scale), $\varepsilon$ (energy scale), and $m$ (mass of the fluid atoms), with the relevant time scale of $\tau = \sqrt{m\sigma^2/\varepsilon}$. Typical numerical values are $\sigma = 0.3\ nm$, $\tau = 2\ ps$, and $\varepsilon = 120 k_B$, where $k_B$ is Boltzmann constant.[68] The time-step in the simulation is selected as $\Delta t = 0.002\tau$. The smallest Janus particle is composed of 88 atoms and is roughly *2 nm* in diameter, while the largest nanoparticle contains 720 atoms and has a diameter of *~4 nm*.

**APPENDIX B. Fluid properties**

The average fluid properties are evaluated in a simulation box containing only two fluid species. The fluid density is measured by decomposing the domain into 250 layers parallel to the interface plane and averaging the number of fluid atoms per unit volume over $10^4$ time steps after equilibration. Figure 10 shows the density profile for each fluid and the total density as a function of the position relative to the interface. The immiscibility of the two fluids is easily confirmed by noting that the number density of each fluid species rapidly approaches zero in the opposite half of the box. Near the interface, there is a total density gap caused by the immiscibility of the two fluids. Such density gap has also been observed in other molecular simulations of fluid interfaces,[69] and is usually not captured in continuum models.[70] Away from the interface, the total density approaches the prescribed value of $\rho = 0.8/\sigma^3$. The calculated average density is within 1.5% of this assigned value. The interfacial width can be estimated from the density profiles of individual components as the distance between the points at which the density of each fluid drops from 90% to 10% of its bulk value.[71] This yields an average width of $w \approx 2.5\sigma$, which translates into *7.5 Å* in metric units and is typical for simple fluid interfaces. The diffusivity of the fluid atoms is determined by calculating the Mean-Squared Displacement (MSD). This requires evaluating the actual displacement of each fluid atom *i* as

$$\Delta r_i(t)^2 = \left(x_i(t) - x_i(0)\right)^2 + \left(y_i(t) - y_i(0)\right)^2 + \left(z_i(t) - z_i(0)\right)^2. \tag{B1}$$

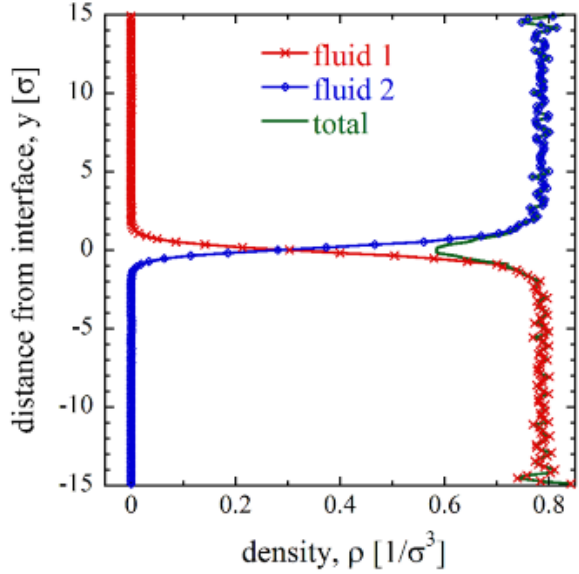

FIG. 10. Density profiles of the two fluids as a function of normal distance from the interface plane.

Averaging the displacements for fluid atoms far from the walls gives the mean-squared displacement versus time, the slope of which yields the fluid diffusion coefficient

$$D = \frac{1}{6}\frac{d}{dt}\langle \Delta r_i(t)^2 \rangle = \frac{1}{6}\frac{d}{dt}\langle (r_i(t) - r_i(0))^2 \rangle. \tag{B2}$$

This was calculated as $D = 0.069\ \sigma^2/\tau$ for the above type of fluid, which is within 1.5% of the value predicted by the available correlations.[72]

To quantify the degree of immiscibility between the two fluids, the interfacial tension calculations are carried out. This quantity is related to the difference between the normal and transverse components of the stress tensor integrated over the longitudinal (y) direction

$$\gamma = \int_{-L_y/2}^{L_y/2} [P_N(y) - P_T(y)]\, dy. \tag{B3}$$

We perform the interfacial tension calculation in a system with similar size but replacing the limiting walls with periodic boundary conditions in y direction. Therefore, two interfaces will exist in the simulated system and the interfacial tension is translated into

$$\gamma = \frac{1}{2}\int_{-L_y/2}^{L_y/2} [P_{yy} - 1/2(P_{xx} + P_{zz})]\, dy. \tag{B4}$$

In the absence of any external constraint, the pressure tensor is expected to be diagonal. Due to the symmetry in our simulation box, two of the components are equal: $P_{xx} = P_{zz}$. Hydrostatic equilibrium imposes the condition that the normal pressure $P_{yy}$ is constant everywhere, while the transverse pressure will differ in the interface region. Using the Kirkwood-Buff formulation of the tangential and normal components of the pressure tensor, the interfacial tension can be written as[69]

$$\gamma = \frac{1}{4A}\langle \sum_{i<j} \left(1 - \frac{3y_{ij}^2}{r_{ij}^2}\right) r_{ij}\, u'(r_{ij}) \rangle, \tag{B5}$$

where $A$ is the interfacial area ($L_x \times L_z$), $r_{ij}$ is the distance between two atoms $i, j$, $y_{ij}$ is their distance in y direction, and $u'(r_{ij})$ is the spatial derivative of the potential function: $\partial u(r_{ij})/$

$\partial r_{ij}$. To calculate the interfacial tension, we sum over all interacting pairs after an initial equilibration period and calculate $\gamma$ by averaging over time and different realizations in Eq. (B5) assuming ergodicity.[65] We obtained $\gamma = 1.47\ \varepsilon/\sigma^2$, which is within 2% of the values reported in similar MD studies.[69] Using the given typical values for $\varepsilon$ and $\sigma$, this translates into *27.0 mN/m*, which is consistent with a typical silicone oil-water interface.

**APPENDIX C. Nanoparticle diffusion in bulk**

In order to validate the molecular simulations with Stokes-Einstein theory, we simulate the diffusion of nanoparticles in bulk by assigning attraction coefficients $A_{12} = A_{21} = 1.0$ between the two fluid species. Homogeneous solid particles are initially placed at the center of the simulation box, and the position, orientation, as well as linear and angular velocities are tracked in multiple realizations over $2\times10^5$ steps after the equilibration stage. Five particle sizes in the range of $3 \leq R/\sigma \leq 6$ are considered. According to Stokes-Einstein theory, the translational diffusion coefficient of a particle with radius $R$ in a bulk fluid of viscosity $\eta$ and no-slip boundary conditions is given by

$$D_t = \frac{k_B T}{6\pi R \eta}. \tag{C1}$$

Similarly, the rotational diffusion coefficient of the particle is given by the Stokes-Einstein-Debye relation

$$D_r = \frac{k_B T}{8\pi R^3 \eta}. \tag{C2}$$

For Lennard-Jones fluids with the assigned temperature and density, the viscosity is determined from the available tables as $\eta \approx 2.0\ m\sigma^{-1}\tau^{-1}$ and used in Eqs. (C1,C2).[64] The translational and rotational diffusion coefficients for nanoparticles of different sizes are plotted in Fig. 11 along with their theoretical values. We observe that the simulated values are in good agreement with the theory, showing typically <6% deviation. The theoretical values lie within the statistical uncertainty of simulated results, thus validating the molecular simulations. The deviation can be due to the inaccuracy in the estimated viscosity, which was obtained by linear interpolation of tabulated values as well as the round-off error in integrating velocity correlation functions. In addition, the simulated particles are not smooth spheres and there is some ambiguity in the definition of particle radius, as reported in similar molecular simulations.[54, 68, 73]

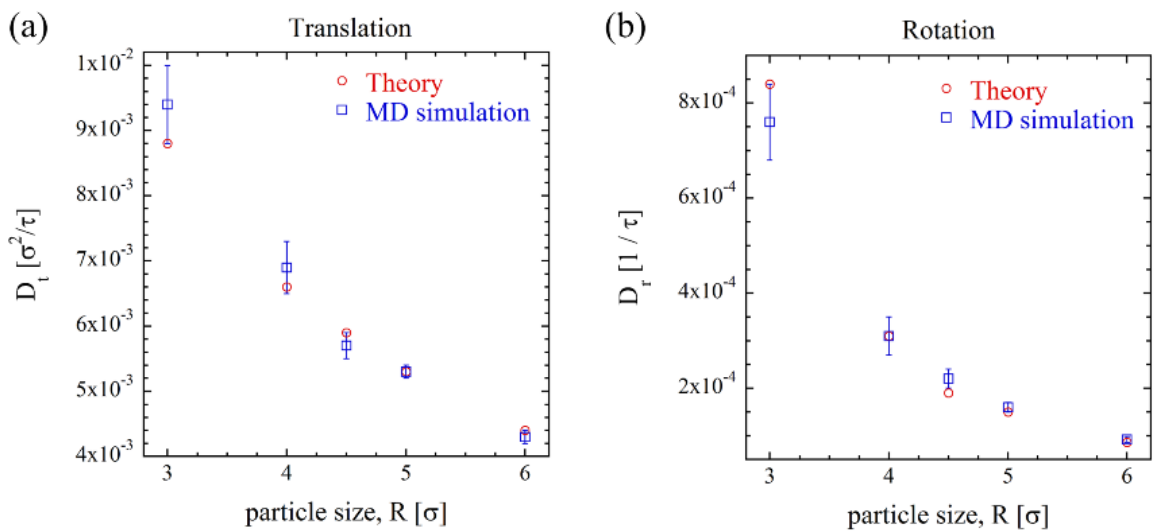

FIG. 11. (a) Translational and, (b) rotational diffusion coefficients for homogeneous nanoparticles of different sizes in a bulk Lennard-Jones fluid and the theoretical values found from Stokes-Einstein(-Debye) relations (Eqs. C1,C2).